# Diffusion Limited Aggregation with modified local rules

Bogdan Ranguelov[1], Desislava Goranova[1], Vesselin Tonchev[1], Rositsa Yakimova[2]
[1]Institute of Physical Chemistry – Bulgarian Academy of Sciences
[2]Linkoping University, Sweden

**Abstract**
Results from a modified Diffusion Limited Aggregation (DLA) model are presented. The modifications of the classical DLA model are in the attachment to the cluster rules and in the scheme of particle generation/killing. In the classical DLA model if a particle reaches the growing cluster it sticks to it immediately and irreversibly and then the next particle is released. We will abandon this original prerequisite, and by changing the sticking probability to the cluster we will change the diffusion regime towards more kinetic one. For a growing cluster consisting of only one type of particles this variation in the sticking probability is (more or less) a rude violation of the hypothesis for diffusion limitation in the DLA model. Since in a lot of experiments different types of particles are used with different sticking probabilities (e.g. different regimes of attachment), we develop a modified DLA model with two types of particles. The second modification we introduce at that point is a scheme for particle generation/killing we call "second chance" – when a particle is killed after reaching a given limiting distance from the cluster, it is killed and then returned to the point it was originally generated. Thus the model is capable to produce a great variety of growing patterns (fractals, spirals) by changing only a single parameter and we are able to construct a morphological diagram of our generalized DLA model with two different types of particles.

**keywords:** aggregation, pattern formation, fractals

**Introduction**

Despite of its simplicity, the Diffusion Limited Aggregation (DLA) model [1] continues to attract attention in different fields such as epitaxial growth [2], snowflakes formation [3], electrochemical deposition [4, 5] and even dielectric breakdown [6]. Common feature for all this processes is that the growth patterns are formed under non-equilibrium conditions and the rate of growth is controlled by material transport via slow diffusion – as a result (generally) a ramified fractal structure is grown and it is characterized with fractal dimension $D_f$. Here we will answer to two questions – what will happen if we start to lower the sticking probability for attachment to the cluster, and what kind of patterns we will obtain if we introduce two different types of particles that are able to stick to the cluster.

**The Model**

Generally our model differs from the classical DLA only by a change in the local attachment rule. While in the classical DLA a particle is immediately attached to the cluster, in our first modification of the model we will introduce the so called sticking coefficient Kstick with ($0 < \text{Kstick} \leq 1$) – it defines the probability for attachment to the cluster. So, in the classical DLA model Kstick equals one, and by lowering this coefficient we will give more and more chances of a particle to diffuse between the branches of the fractal cluster. The "growth procedure" goes like in the classical DLA model: a seed particle (the first one from the cluster) is placed in the centre of a square lattice. The next particle is released from a random place onto a ring with radius Rmax + BR (the so called birth radius) and centered at the seed. Here Rmax is the maximal temporary radius of the cluster and BR is an offset (in the beginning Rmax equals zero). The released particle performs a random walk on the square lattice until it reaches the seed. Then if Kstick is equal to one, the particle is attached to the seed. If Kstick is lower than one, a random number is generated and is compared with the value of Kstick. If the random number is lower than Kstick, then the particle is attached to the seed, otherwise the random walk process continues until the seed is reached again and so on. When the second particle is attached (sooner or later) to the cluster then next particle is generated and released, the random walk and attachment procedures are same for all particles and thus the cluster grows by adding particles one by one.

**Results I**

Typical cluster shapes for four different values of Kstick are shown on Fig.1. It is clearly seen that lowering the value of Kstick leads to a more and more dense packing structure. The reason is that the lower Kstick diminishes the capture "activity" of the outermost branches of the cluster and allows a particle to diffuse towards the center and thus to fill the empty gaps onto the lattice.

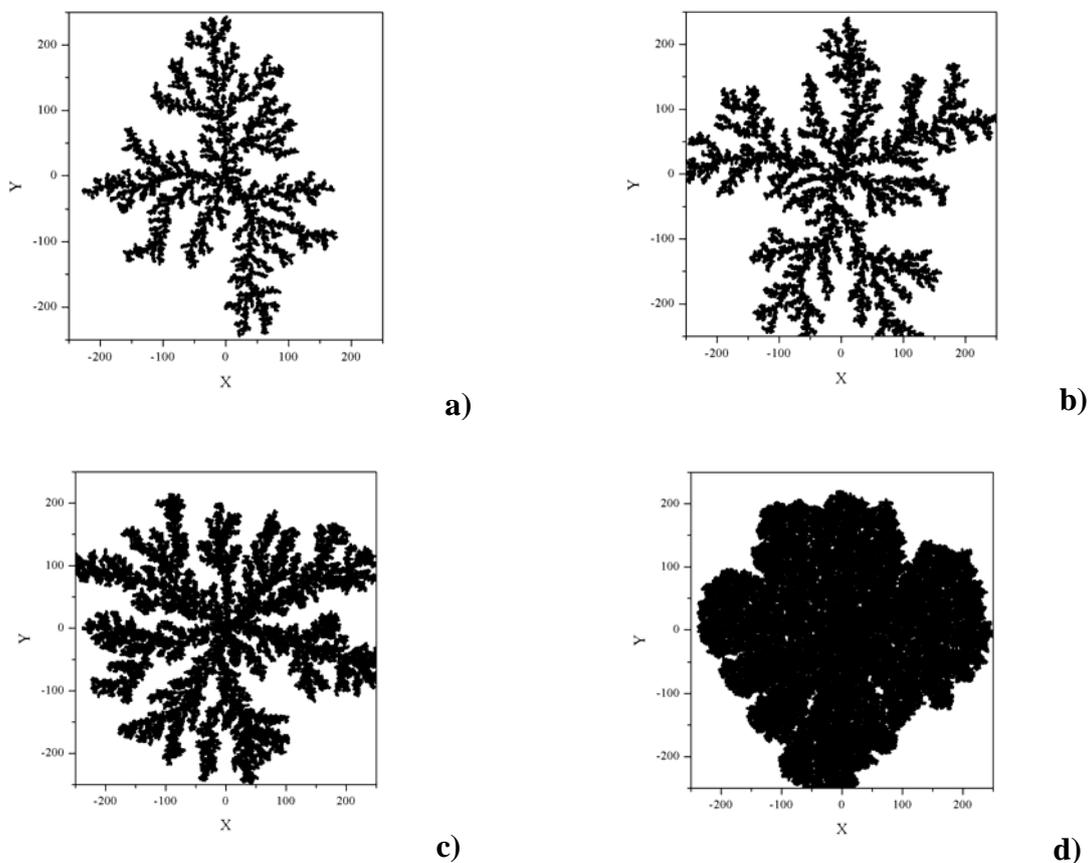

a) b) c) d)

Fig.1. Four growing DLA clusters with different sticking probability: Kstick = 1.00; 0.50; 0.10 and 0.01. For all clusters BR = 100.

This is a morphology transition due to the lowering of the sticking coefficient, which is seen also by the change in the fractal dimension $D_f$ of the clusters – see Fig.2. The fractal dimension relates the number of particles N attached to the cluster with the maximum radius of the cluster, so $N \sim R_{max}^{D_f}$. BR is chosen to be 100, since lower values lead to a generic instability that leads to a quasi one dimensional growth [6].

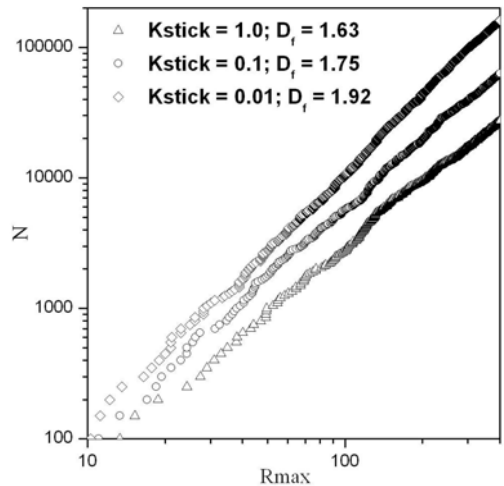

Fig.2. Fractal dimension of the growing cluster for different values of the sticking coefficient: 1.0, 0.1 and 0.01. For all clusters BR = 100.

**Two particles DLA model**

     We modify the classical DLA model by concerning two different types of particles that are released and subsequently aggregate to the cluster. In comparison with the classical DLA model now we have BR1 and BR2 (birth radii), already introduced in the beginning. Here 1 stands for particles type1 and 2 for particles type2 respectively. In addition introducing two types of particles leads (generally) to four different sticking coefficients, which we denote by Kstick11, Kstick22, Kstick12 and Kstick21. This sticking coefficients represent the probability for sticking between particles of same type and different types – for instance K12 is the probability for sticking particle type1 on particle type2, etc. We will call K11 and K22 homo-sticking coefficients, while K12 and K21 hetero-sticking coefficients respectively. Thus by varying homo and hetero sticking coefficients we are able to distinguish different regimes of attachment for the two different types of particles. Here we will demonstrate results that are obtained with zero hetero-sticking coefficients. The initial seed consists of one particle of type1 and one particle of type2. Now we will propose one main and important difference in comparison with all other DLA models – we will release the particles not from ring with a radius Rmax + BR, but from circumference at distance BR from the real circumference of the growing cluster – see Fig.3. A simple reason for this is the experimentally observed density profiles in the solution [8] around the cluster in a quasi 2D electrochemical cell. Defining this release profile is done by probing the cluster surface with dummy particles of different sizes. Similarly one can define another circumference at distance KR from the real circumference of the cluster and KR>BR. KR represents the surface beyond which one can consider that the particle is too far from the cluster and it will never stick to it. Hence, this particle is "killed" and another one is generated (similar procedure is used in [9]). This procedure saves not only computer time, but has a clear physical meaning – if a particle is too far from one cluster the chance to "come back" and to stick is negligible, so we can think about this particle as a "lost one". The second main difference in comparison with all other DLA models is the condition that if a particle crosses the KR "line" then it is returned at the same place where it was initially generated.

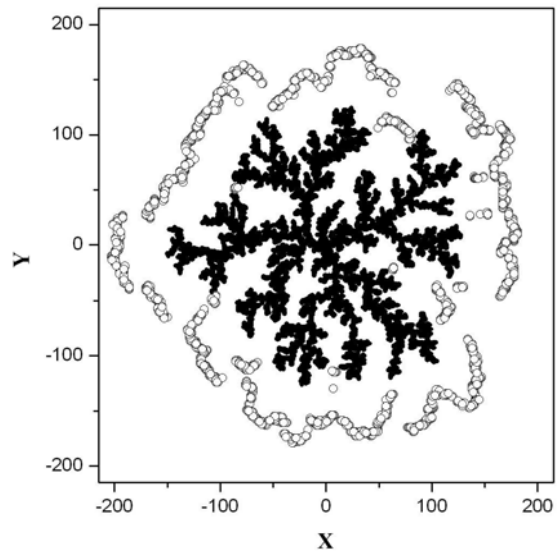

Fig.3. A cluster (black dots) and the corresponding circumference at distance BR = 50 (open circles) from the real cluster circumference.

**Results II**

Here we will show results for equal concentrations of particles type1 and type2 and zero hetero sticking coefficients – see Fig.4.

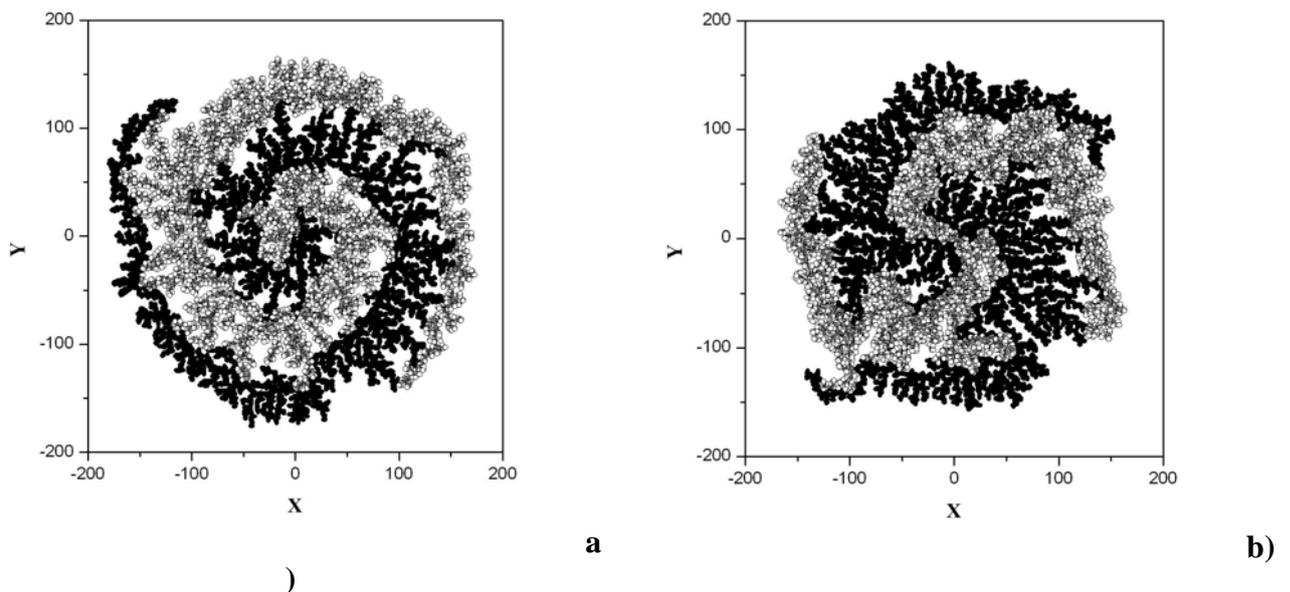

Fig.4. Particles type1 – black dots, particles type2 – open circles: **a)** BR = 5; KR = 30; Kstick11 = Kstick22 = 1.0; **b)** BR = 5; KR = 30; Kstick11 = 1, Kstick22 = 0.1

Lower values of BR produce spiral patterns in which every single spiral arm consists of one type of particles. Depending on the initial size of the seed we are able to produce single to many arm spirals, rotating both in clock and clockwise directions. Closer look into the structure reveals that the intrinsic structure of the spiral arm is fractal (which is governed by the value of the homo sticking coefficients) – see Fig.4a. Lower values of Kstick11 and/or Kstick22 can produce dense compact structures or dense structure from one type of particles together with fractal like structure from the other type of particles – see Fig.4b.

The influence of the BR value is illustrated no Fig.5. There is a critical value around BR=100 around which the morphology of the cluster is drastically changed from a spiral to a ramified fractal one.

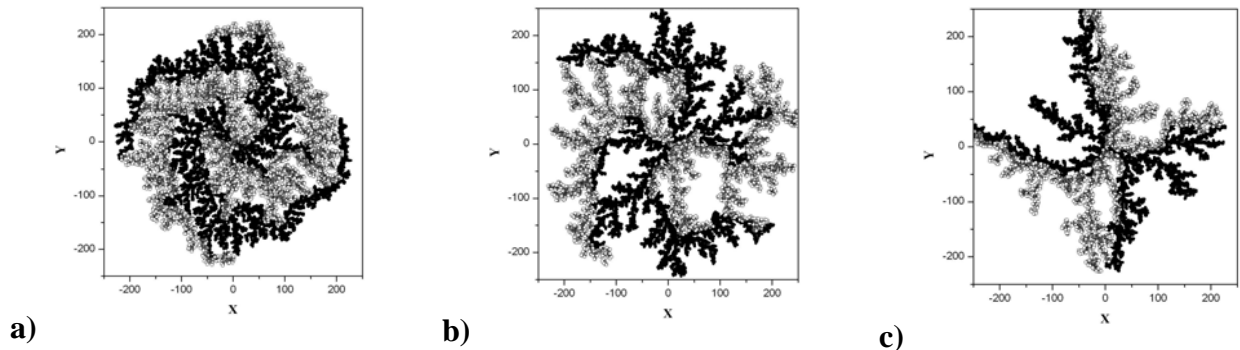

Fig.5. Particles type1 – black dots, particles type2 – open circles: **a)** BR = 10; KR = 40; Kstick11 = Kstick22 = 1.0; **b)** BR = 80; KR = 110; Kstick11 = Kstick22 = 1.0 **c)** BR = 200; KR = 240; Kstick11 = Kstick22 = 1.0

Here we show results only for the case of equal concentrations, equal "birth radii" and zero hetero sticking coefficients. By changing a single parameter in our model we are able to construct wide variety of cluster morphologies emerging in various systems driven far from equilibrium.

**Summary**

In the framework of a single model based on a classical DLA procedure we are able to produce not only ramified fractal structures, but also spirals and "palm leaf" [10] patterns that are typical for systems with different interactions (electro co-deposition [11], chemical reactions on surfaces [12], or cracking in sol-gel films [13] ) that are driven far from equilibrium.

**Acknowledgements:**

The authors acknowledge the stimulating working conditions by participating in Projects MADARA and IRC-CoSIM funded by National Science Fund - Bulgaria. One of the authors (RY) is grateful for the financial support by the Swedish Research Council.